\documentstyle[epsf]{article}
\begin{document}
\begin{center}
{\Large \bf Dileptons as Probes of \\
            High-Density Hadronic Matter: \\[.2cm]
            Results from the SPS Heavy-Ion Programme}
\end{center}
\vspace{0.3cm}

\begin{center}
      {Itzhak Tserruya \footnote{Work supported by research grants from  the Israel Science Foundation
                                 and the Fusfeld Research Fund.}}\\
      \vspace{0.3cm}
      {\it Weizmann Institute of Science, Rehovot, Israel}
\end{center}
\vspace{0.5cm}

{\bf Abstract}
The most recent results on dileptons obtained 
in the CERN heavy-ion programme are reviewed. The emphasis is on the excess of 
low-mass lepton pairs observed in the CERES, HELIOS-3 and NA38/50 experiments
which seems to point at modifications of the vector meson properties, and in
particular the $\rho$ meson, in a high density baryonic medium. Recent results
on intermediate mass dileptons  are also presented. 
\vspace{1.0cm}
 
\section{Introduction}

  The heavy-ion programme at the CERN SPS started in 1986 with the acceleration of 
O beams at an energy of 200 GeV/c per nucleon followed soon after 
by a S beam at the same energy. Since 1994 the programme uses a Pb beam of 
158 GeV/c per nucleon. Among the vast amount of
experimental results that have been gathered 
%covering a large variety of 
%experimental variables, a significant fraction of them hinting at new physics. 
the observation of an excess emission of low-mass dilepton pairs appears as one of the most 
notable and intriguing achievements of the programme along with the J/$\psi$ suppression
and strangeness enhancement \cite{qm}.
   
Dileptons emitted in ultra-relativistic heavy-ion collisions are considered 
unique probes in the study of hadronic matter under extreme conditions of temperature 
and baryon density and in particular the conjectured deconfinement and chiral phase 
transitions. These penetrating probes have a relatively large mean free path and consequently
can leave the interaction region without final state interaction, carrying information about 
the conditions and properties of the matter at the time of their production
and in particular of the early stages of the collision when temperature and energy density have 
their largest values 
\footnote{The same argument is in principle valid for real photons, since real and 
virtual (dileptons) photons are expected to carry the same physics information. However, 
the physics background for real photons is larger by orders of magnitude as compared 
to dileptons, making the measurement of photons much less sensitive to a new source.}.
%This has to be contrasted with the hadronic observables 
%which are sensitive to the late stages of the collisions at freeze-out, 
%i.e. once the hadronic system stops interacting.
  
   A prominent topic of interest is the identification of thermal radiation emitted from 
the collision system. This radiation should tell us about the nature of the matter formed, 
a quark-gluon plasma (QGP) or a high-density hadron gas (HG). The elementary 
processes involved are well known: {\it q}{\it $\overline{q}$\/} annihilation 
in the QGP phase and $\pi^+\pi^-$ annihilation in the HG phase and the two sources 
should be distinguishable by their characteristic properties in appropriate mass windows. 
At SPS energies, the initial temperature is believed to be close to the critical 
temperature, and therefore one expects the dense hadron gas to be the dominant source 
of thermal radiation. The window to search for it  is at low masses, around and below 
the $\rho$-meson mass \cite{kajantie86,cleymans91-koch93}, since the $\pi^+\pi^-$ 
annihilation cross section is dominated by the pole of the pion electromagnetic form 
factor at the $\rho$ mass. On the other hand, theoretical calculations have singled out 
the mass range of 1-3 GeV/c$^2$ as the most suitable window to observe the thermal 
radiation from the QGP phase~\cite{kajantie86,ruuskanen92} at initial temperatures 
likely to be reached at RHIC or LHC.

The physics potential of dileptons is further emphasized by the capability to measure 
the vector mesons  \cite{shor-phi-heinz} through their leptonic decays. Of particular 
interest is the decay of the $\rho$ meson into a lepton pair since it provides a unique 
opportunity to observe in-medium modifications of the vector meson properties
which might be linked to chiral symmetry restoration. 
Due to its very short lifetime  ($\tau$ = 1.3 fm/c) compared to the typical fireball 
lifetime of 10-20 fm/c at SPS energies, most of the $\rho$ mesons produced in the 
collision will decay inside the interaction region with  modified mass and/or width 
if the temperature or the baryon density are large enough. The relation of in-medium 
modifications to chiral symmetry restoration is a highly debated topic and will be 
addressed again in this paper \cite{rapp-qm99}.   
The situation is very different for the other vector mesons, $\omega,\phi$ or $J/\psi$. Because 
of their much longer lifetimes they will be reabsorbed in the medium or they will decay 
well outside the interaction region after having regained their vacuum masses.

  The three experiments, CERES, HELIOS1-3 and NA38/50 involved 
in the measurement of dileptons ($e^+e^-$ and $\mu^+\mu^-$ pairs) at the CERN SPS have 
indeed confirmed the unique physics potential of these probes.
   The results cover measurements with p, S and Pb beams, and the most notable one
is the observation with the ion beams of an enhancement of the 
dilepton yield over a very broad range of invariant masses. The enhancement is particularly 
pronounced
in the continuum at low-masses (0.2 $<$ m $<$ 0.7 GeV/c$^2$) but it is also significant in the 
continuum at intermediate masses (1.5 $<$ m $<$ 3.0 GeV/c$^2$) and in the $\phi$ meson yield.
The low-mass pair enhancement has triggered a huge amount of theoretical activity mainly
stimulated by interpretations based on in-medium modifications of the vector mesons and in 
particular a decrease of the $\rho$-meson mass as a precursor of chiral symmetry restoration.

\begin{table}[h!]
\vspace{-0.5cm} 
\begin{center}
{ Table 1. List of Dilepton Measurements at the CERN SPS} \\[0.3cm]
\leavevmode
\begin{tabular}{||c|c| c| c|  c   |c ||}\hline\hline

 Experiment &  Probe      &  System           &  $y$      &    Mass       & Ref. \\
            &             &                   &           & (GeV/c$^2$)   &      \\ \hline
            &             & p-Be,Au 450 GeV/c &           &               & 7,8  \\ 
 CERES      &   $e^+e^-$  & S-Au    200 GeV/u &  2.1-2.65 & 0 -- 1.4      & 9    \\
            &             & Pb-Au   158 GeV/u &           &               & 10-12  \\[0.2cm]
  \hline 
 HELIOS-1   &$\mu^+\mu^-$ & p-Be 450 GeV/c    & 3.65-4.9  & 0.3 -- 4.0    & 13    \\
 (completed)& $e^+e^-$    &     ``            & 3.15-4.65 &      ``       & 13   \\[0.2cm]
\hline
 HELIOS-3   &$\mu^+\mu^-$ & p-W,S-W 200 GeV/u & $>3.5$    & 0.3 -- 4.0    & 14   \\
 (completed)&             &                   &           &               &      \\[0.2cm]
\hline
 NA38       & $\mu^+\mu^-$& p-A,S-U 200 GeV/u & 3.0-4.0   & 0.3 -- 6.     & 15   \\
 NA50       &     ``      & Pb-Pb   158 GeV/u &    ``     & 0.3 -- 7.0    & 16   \\[0.2cm]
\hline
\hline
\end{tabular}
\end{center}
\end{table}

\vspace{-0.5cm}
 A compilation of all measurements performed so far together with the kinematic phase space 
covered and relevant references is presented in Table ~1. 

   In this paper, I concentrate on the most recent experimental results on low-mass lepton 
pairs (Section2) and current status of the interpretations (Section 3). Other recent reviews can be
found in \cite{it-hiroshima}. Section 4 deals with 
the intermediate mass region and Section 5 gives some concluding remarks and prospects.

\section{Low-mass Dileptons: Experimental Results}

  The low-mass region, m = 200 - 600 MeV/c$^2$, has been systematically studied by the CERES 
experiment \cite{pbe-ee,pbe-eegamma,prl95,ir-qm97,pbau95-plb,bl-qm99}. The most
recent results were obtained with the Pb beam in two different runs, in 1995
\cite{ir-qm97,pbau95-plb} and 1996 \cite{bl-qm99}.
Apart from a slight difference in the centrality trigger  ($<$dn$_{ch}$/d$\eta>$ = 250 and 220
in the '96 and '95 runs,  respectively), the two measurements were performed under identical
conditions. The invariant mass spectra from the two data sets, normalized to the measured 
charged particle rapidity density are displayed in Fig.1.   
 On the positive side, one notes that the results appear consistent with each other 
within their error bars. 
The level of agreement is remarkable if one keeps in mind the huge filtering 
of the data 
\footnote {Figure 1 contains a total of 2018 (648) $e^+e^-$ pairs with mass m $>$ 200 MeV/c$^2$ 
out of 42.2x10$^6$ (8.6x10$^6$) analyzed events in the '96 (95) run.},
the fact that these are two different data sets, and that they have been
analyzed with the same strategy but with a somewhat different technique  
\footnote {
However, one also immediately notices that the '96 results are systematically lower.
The origin of the effect was studied \cite{socol-thesis} and traced back to
subtle differences in the analysis procedure.
In '96  all cuts were tuned without making use of the data themselves,
basing  them for example on Monte Carlo simulations \cite{bl-qm99, socol-thesis,lenkeit-thesis}. 
The '95 data analyzed with the same Monte Carlo tools as in '96 yield nice 
agreement between the two data sets within statistical errors. The differences is Fig.1 are
therefore a reflection of the systematic errors of the two analysis methods.}.
 
%%%%FIGURE 1. CERES 95-96 RESULTS WITH PP AHD THERMAL COCKTAILS 
\begin{figure}[h!]
\vspace{-1.0cm}
\centerline{\epsfxsize=8.5cm \epsfbox{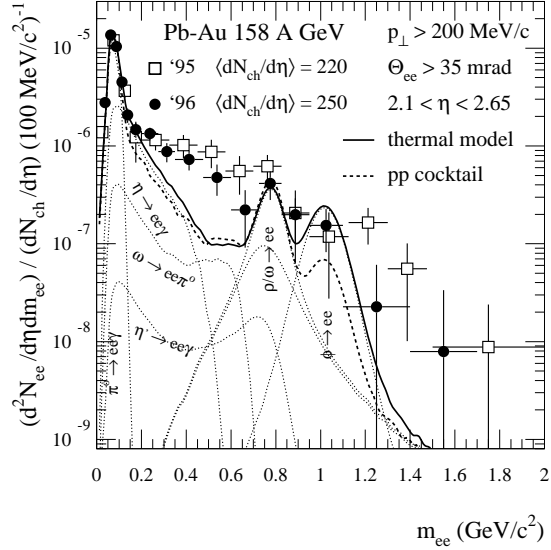}}
\vspace{-0.7cm}
\caption {Inclusive $e^+e^-$ mass spectrum measured by CERES in 158 A~GeV Pb--Au 
         collisions in the '95 and '96 runs.  The figure also shows the summed (solid line) and 
         individual (dotted lines) contributions from hadronic sources in a thermal model 
         ~\protect\cite{bl-qm99}. The predictions from the pp cocktail 
         previously used by CERES \cite{pbe-ee} are shown by the dashed line.}
\end{figure}

%THE COKTAILS AND THE ENHANCEMENT

   As previously observed with the S beam \cite{prl95}, the $e^+e^-$ pair yield is clearly enhanced 
in the mass range above $\sim 200 MeV/c^2$  and below the $\rho/\omega$ peak, 
with respect to the expected yield from known hadronic sources.
The solid line shows the total expected yield based on a generator \cite{bl-qm99}
which uses 
measured particle production ratios whenever available or ratios calculated with 
a thermal model which describes well all these ratios \cite{pbm-js}. With respect to this 
cocktail the measured yield in the mass region m = 0.25 - 0.7 GeV/c$^2$ is enhanced by a factor
of 2.6 $\pm$0.5(stat.) $\pm$0.6(syst.). For comparison the figure also shows 
(dashed line) the standard pp cocktail previously used in the 
presentation of the CERES results \cite{pbe-ee} and which is based on yields directly measured in pp
collisions, scaled to the nuclear case with the charged particle rapidity density. 
The two generators predict closely similar results. The total yield of the thermal model is
$\sim$30 \% larger than the pp cocktail for masses m $>$ 200 MeV/c$^2$, the main difference occuring 
in the region of the $\phi$ meson.

   CERES has further characterized the properties of the low-mass excess by studying its p$_t$ and
multiplicity dependences, which indicate that the excess is mainly due to soft pair p$_t$ and increases
faster than linearly with the charged particle density 
\cite{ir-qm97,pbau95-plb,bl-qm99}. 

%HELIOS3 AND NA38 RESULTS - FIGURE OF NA38.
%%%%%   NA50 results
   An enhancement of low-mass dileptons has also been observed in the di-muon experiments
HELIOS-3  \cite{helios-3} and NA38 \cite{falco-qm97} with the S beam. NA38 has an interesting set of
results including p-U, S-S and S-U collisions at 200 A GeV. Whereas the p-U data are well reproduced
by a cocktail of hadronic sources (with the somewhat uncertain extrapolation of the 
Drell-Yan contribution into low masses), the S data shows an enhancement of low-mass pairs. 
The enhancement is most apparent in the S-U collision system and there it clearly
extends over the intermediate mass region as illustrated in Fig.2.
%%%%%% FIGURE 2. NA38 S-U  MASS SPECTRUM 
\begin{figure}[h!]
\vspace{-0.3cm}
\centerline{\epsfxsize=8cm \epsfbox{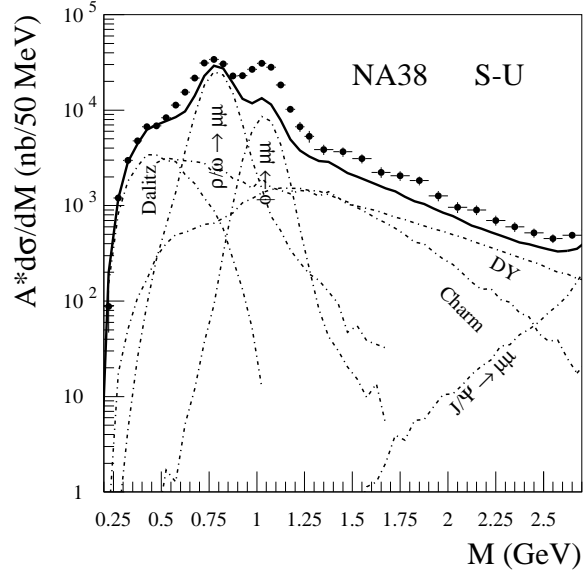}}
\caption{Inclusive $\mu^+\mu^-$ mass spectra measured by NA38 in 
              200 A~GeV S-U collisions. The thick line represents the summed 
              yield  of all known sources. The individual contributions
              are also shown ~\protect\cite{falco-qm97}.}
\label{fig:na38-su}
%\vspace{-0.9cm}
\end{figure}

There is a striking difference in the shape of the low-mass dilepton spectrum as 
measured by CERES and NA38. A pronounced  structure due to the resonance decays 
is clearly visible in the NA38 spectrum, whereas in the CERES results the 
structure is completely washed out (see Fig. 1), raising the 
question of consistency between the two experiments. Resolution effects can 
be readily ruled out since the low-mass spectrum in p-Be and p-Au collisions 
measured by CERES with the same apparatus clearly shows the $\rho$/$\omega$ peak 
\cite{pbe-ee}. We also note that the two experiments cover nearly symmetric ranges 
around mid-rapidity ($\eta$ = 2.1 -- 2.65 and $\eta$ = 3 -- 4 in CERES and NA38 
respectively). But  CERES has a relatively 
low $p_t$ cut of 200 MeV/c on each track whereas NA38 is restricted to  
$m_t > $ 0.9 + 2($y_{lab} - 3.55)^2$ GeV/c$^2$. Moreover, NA38 has no centrality selection 
in the trigger whereas the CERES data corresponds to the top 30\% of the geometrical cross 
section. These two factors are likely to explain the apparent discrepancy since, as noted previously,
the excess observed by CERES is more pronounced at low pair p$_t$ and increases stronger than 
linearly with multiplicity.
Given enough statistics it should be fairly easy for the two experiments 
to apply common $m_t$ and centrality  cuts thereby  making possible a 
direct and meaningful comparison between their results.
 
\section{Low-mass Dileptons: Theoretical Evaluation}
 
   The enhancement of low-mass dileptons has triggered a
wealth of theoretical activity. Dozens of articles have been published on the subject and clearly
it is not possible to review them here. I present thus a summary of the
current leading approaches.  
 There is a consensus that an 
additional source beyond a simple superposition of pp collisions is needed. Furthermore, it is commonly 
recognized that the pion annihilation channel ($\pi^+\pi^- \rightarrow l^+l^-$), 
obviously not present in pp collisions, has to be taken into account. This channel 
accounts for a large fraction of the observed enhancement 
%(the  thin line in Figure ... contains the pion annihilation 
%in addition to the hadron decays) 
however it is not sufficient to reproduce the data in the mass region  
0.2 $< m_{e^+e^-} <$ 0.5 GeV/c$^2$. These data have been quantitatively 
explained by taking into account in-medium modifications of the vector mesons.
Li, Ko and Brown \cite{li-ko-brown} were the first to propose and use a decrease of 
the $\rho$-meson mass in the hot and dense fireball as a precursor of chiral 
symmetry restoration, following the original Brown-Rho scaling \cite{brown-rho}. With this approach,  
an excellent agreement with the CERES dat is achieved  
as demonstrated by the solid line in Fig.3 (taken from \cite{rapp-qm99}). 
%%% FIGURE 3 CERES 96 PB RESULTS WITH BR, RAPP AND KOCH. 
\begin{figure}[h!]
\vspace{-1.0cm}
\centerline{\epsfxsize=8cm \epsfbox{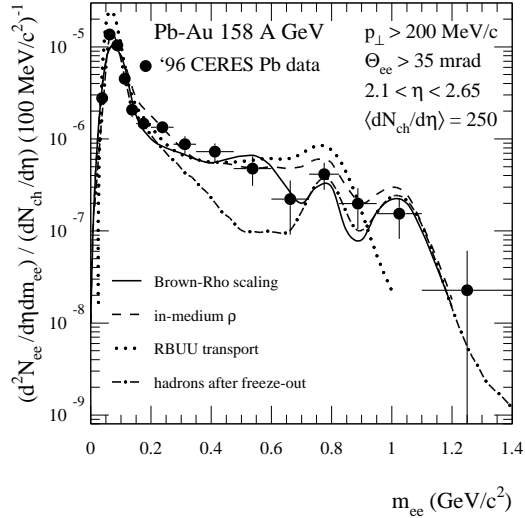}}
\vspace{-0.8cm}
\caption {CERES results
         compared to calculations using dropping $\rho$ mass 
         (Brown-Rho scaling), in-medium $\rho$-meson broadening and RBUU transport model. 
         The dash-dotted line represents the yield from hadrons after freeze-out as in Figure 1.   }
%\vspace{-0.5cm}
\end{figure}
 
 Another avenue based on effective Lagrangians uses a $\rho$-meson 
spectral function which takes into account the $\rho$ propagation in hot and dense
matter, including in particular the pion modification in the nuclear medium and 
the scattering of $\rho$ mesons off baryons \cite{wambach, cassing-wambach}. This leads to a large 
broadening of the $\rho$-meson line shape and consequently to a considerable 
enhancement of low-mass dileptons. These calculations achieve also an excellent reproduction
of the CERES  results as illustrated by the dashed line in Fig. 3.
Although the two approaches are different in the underlying physical picture 
(in the Brown-Rho scaling the constituent quarks are the relevant degrees of freedom 
whereas ref. \cite{wambach} relies on a hadronic description), it turns out that
the dilepton production rates calculated via hadronic and partonic models 
are very similar at SPS conditions \cite{rapp-qm99} thus explaining the similar results
of the two approaches.
Several issues remain controversial.  
Both approaches rely on a high baryon density for the dropping mass or the
enlarged width of the $\rho$ meson but the role of baryons is still a question open to
debate. Calculations based on chiral reduction formulae, although similar in principle
to those of ref. \cite{wambach}, find very little effect due to baryons and are in fact 
low compared to the data \cite{zahed}.  The RBUU transport
calculations of Koch \cite{koch} find also very little effect due to the baryons and come to a 
reasonably close description of the data as shown in Fig. 3 by the dash-dotted line. 
This could be due to an  overestimation of the  $\omega$ Dalitz decay yield
as a consequence of an increased  $\omega$ yield directly reflected in the figure 
at m $\sim$ 800 MeV, which is dominated by the  $\omega \rightarrow e^+e^-$
decay. Finally, I wish to point out the discrepancy between transport \cite{cassing-wambach}
 and hydrodynamic calculations \cite{prakash} in treating the time evolution of the fireball,  
the former yielding a factor of 2-3 higher yields.

\section{Intermediate-mass Dileptons}
   The results of HELIOS-3 \cite{helios-3} and NA38/50  \cite{falco-qm97, bordalo-qm99} 
clearly show an  excess of dileptons in the 
intermediate mass region 1.5 $<$ m $<$ 3.0 GeV/c$^2$ (see Figs. 2 and 4). The excess refers 
to the expected yield from
Drell-Yan and semi-leptonic charm decay which are the two main contributions 
in this mass region. The shape of the excess is very similar to the open charm contribution
and in fact doubling the latter nicely accounts for the excess. This is the basis for the
hypothesis of enhanced charm production made by NA38/50 \cite{bordalo-qm99}. 
However it is very unlikely that at the
SPS energies charm production could be enhanced by such a large factor \cite{shor}. HELIOS-3
points into a different direction. The excess plotted as a function of the dimuon transverse
mass can be fitted by a single exponential shape below and above the vector mesons \cite{helios-3}, 
suggesting a common origin of the excess in the low and intermediate mass
regions. Following this line, Li and Gale \cite{li-gale}
calculated the invariant dimuon spectrum in 
central S-W collisions at 200 A GeV. 
%%%%%% FIGURE 4. INTERMEDIATE MASS LI CALCULATIONS AND DATA OF HELIOS 3
\begin{figure}[h!]
\vspace{-0.2cm}
%\begin{minipage}[t]{75mm}
%\framebox[79mm]{\rule[-26mm]{0mm}{52mm}}
\centerline{\epsfxsize=6cm \epsfbox{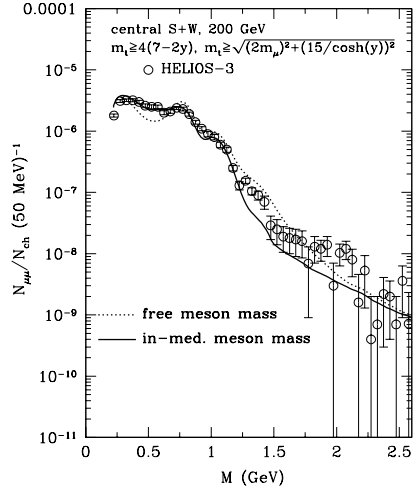}}
\vspace{-0.5cm}
\caption{HELIOS-3 dimuon data compared to calculations with free and in-medium meson masses 
         ~\protect\cite{li-gale}.}
%\end{minipage}
\end{figure}
%\hspace{\fill}
%\vspace{-0.3cm}
On top of the {\it physics}
background of Drell-Yan and open charm pairs, they considered the thermal radiation of muon pairs
resulting from secondary meson interactions including higher resonances and in particular 
the $\pi a_1 \rightarrow l^+l^-$. The calculations are based on the same relativistic 
%Additional contributions like secondary Drell Yan \cite{frankfurt} should 
%be taken into account. 
fireball model used to calculate the low-mass dileptons discussed in the previous 
section \cite{li-ko-brown}. Their results are presented in Fig. 4 showing
the total yield (physics background + thermal yield) with the assumption
of free masses (dotted line) and dropping vector meson masses (solid line).
The latter leads to a much better agreement with the data at low
masses (from 0.3 to 0.7 GeV/c$^2$), as already mentioned in the previous section,
whereas in the intermediate mass region the difference between free and in-medium
meson masses with respect to the data is not so large. The calculations with
free masses slightly overestimate the data whereas with dropping masses 
the situation is reversed.  The intermediate mass region alone cannot be used to validate
the dropping mass model, however it is important that the model can explain
simultaneously the low and intermediate mass regions.

\section{Summary and Outlook}
  The measurements of dileptons both at low and intermediate masses have provided
very intriguing results.
 The outstanding physics question is to further elucidate the origin of the observed excess
and its possible relation to chiral symmetry restoration. 
   There are a number of open questions on the theoretical front: the 
role of baryons, the difference between transport and hydrodynamic calculations, the
approach to chiral restoration (do masses drop to zero and/or do their width increase to infinity?),
With the present accuracy of the data 
it is not possible to discriminate between the 
various models.

 Major new steps are foreseen in the near future. First, 
 CERES is planning to dramatically improve the
mass resolution to achieve $\delta m/m $= 1\%,  by the addition of a TPC 
downstream of the present double RICH spectrometer. With this resolution, 
which is of the order of the natural line width of the $\omega$ meson,
it should be possible to  directly measure the yield of all three vector 
mesons $\rho, \omega$ and $\phi$ including any possible changes in their 
properties (mass shift or increased width) thereby providing a better experimental tool
to reveal possible in-medium modifications of the vector mesons. Second,
a measurement is proposed at the lowest energy attainable at the SPS, at about
40 GeV/nucleon, where the effect of baryon density on the vector meson masses 
is expected to be largest. Last but not least, RHIC start of operations is behind the corner
offering the possibility to extend  these studies under better conditions of energy density and
lifetime and to explore a new domain where temperature  rather than baryon density is expected 
to be the dominant factor. 

%The extension of  the NA measurements to the Pb beam by the NA50 collaboration are very much awaited.
%Results from the

\end{document}